\documentclass{article}
\usepackage{spconf,amsmath,amssymb,graphicx,booktabs,multirow}
\title{Centering Drives Normalization Gains:\\
       Price-Offset Nuisances in Cross-Sectional Return Prediction}

\name{Mingju Chen\textsuperscript{1} \quad
        Qianhui Liu\textsuperscript{2} \quad
        Yui Lo\textsuperscript{3} \quad
        Yuanhang Liu\textsuperscript{4,5,*}\thanks{Corresponding author.} } 

\address{\textsuperscript{1}Shanghai University, Shanghai, China\\
           \textsuperscript{2}School of Artificial Intelligence, University of Chinese Academy of Sciences, Beijing, China\\
           \textsuperscript{3}The University of Sydney, Australia\\
           \textsuperscript{4}Qiuzhen College, Tsinghua University, Beijing, China\\
           \textsuperscript{5}Institute for AI Industry Research (AIR), Tsinghua University, Beijing, China}

\begin{document}
\maketitle

\begin{abstract}
Cross-sectional return prediction from raw intraday bars is sensitive to each
instrument price level, an additive nuisance under a return-ranking hypothesis.
We test whether removing this offset, rather than rescaling amplitudes or
changing the encoder, explains gains on a point-in-time CSI~300 five-minute
panel. We evaluate eight parameter-matched encoders with and without RevIN
normalization; a parameter-free ladder then separates identity, scale-only,
centering, last-value referencing, differencing, and standardization across
all fields and restricted channels. Centering drives the reliable effect,
while scale-only normalization does not help. All eight paired effects are
positive and survive Holm correction on raw rank IC, after style
residualization, and after further residualizing on short-term
reversal. Among six stronger encoders, gains of $0.0376$--$0.0567$ exceed
the $0.0109$ spread of normalized IC ($0.0830$--$0.0939$). Price-only
standardization retains $93$--$101\%$ of the all-field gain. These results
place the main effect in transformed price-channel offset removal rather than
amplitude scaling or encoder choice.
\end{abstract}


\section{Introduction}
\label{sec:intro}

\begin{figure}[t]
\centering
\includegraphics[width=\columnwidth]{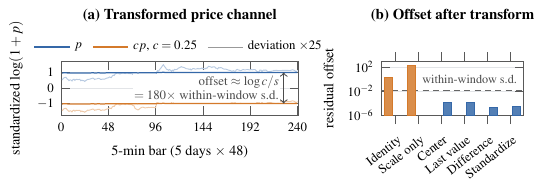}
\caption{Price-level nuisance. (a)~Scaling a path from $p$ to $cp$ preserves
returns but shifts the transformed channel by $\approx\log c$, here about 180
within-window s.d.; exact for $\log p$, approximate for $\log(1+p)$.
Deviations from each mean are magnified $\times 25$. (b)~Offset left by each
transform of Sec.~\ref{ssec:ladder}, in units of the transformed channel's
within-window s.d.; the dashed line marks one s.d.}
\label{fig:motivation}
\end{figure}

Cross-sectional return prediction assigns scores to a stock universe each day
and evaluates their correlation with realized returns. Its long, low-SNR
price--volume input contains a nuisance absent from ordinary forecasting:
multiplying an instrument's price path by a positive constant leaves its return
label unchanged. Under a ranking hypothesis, the corresponding
instrument-specific offsets in transformed price channels should therefore be
removed before prediction. The correspondence is exact after $\log x$ and
approximate under our $\log(1+x)$ preprocessing (Sec.~\ref{sec:problem}). We
test this hypothesis rather than presume that raw price carries no information;
price can also proxy for tick size, liquidity, or investor clientele.
Figure~\ref{fig:motivation} illustrates the nuisance.

Recent cross-sectional models combine latent factors, market-conditioned
attention, state-space blocks, cross-asset graphs, and stock-specific
pre-training
\cite{duan2022factorvae,li2024master,gu2023mamba,mehrabian2025samba,wang2025sspt}. Their
inputs are commonly ratio-style features normalized against recent prices, so
the encoder never sees price level. Architecture, input representation, and
training often change together, making their effects hard to separate. An
analogous confound appears in long-horizon forecasting, where a linear model
becomes competitive under controlled preprocessing
\cite{zeng2023dlinear,toner2024linear}. Sample-level normalization has been
studied for non-stationary forecasting
\cite{kim2022revin,liu2022nonstationary,liu2023san,han2024sin,ye2024fan} and
learned input normalization for financial series
\cite{passalis2020dain,tran2021bin}, but
not the offset-versus-scale mechanism in raw-bar cross-sectional ranking.

To isolate this mechanism we fix one protocol and vary one factor at a time:
eight capacity-matched encoders with and without instance normalization; on
two fixed backbones, removal of temporal offset, temporal scale, or both; and
channel restrictions that attribute the gain to the four price fields or to
volume and traded amount. We contribute (i) a formulation of price level as
an additive-offset nuisance in the transformed input space of return ranking,
with the invariance hypothesis stated precisely, including where the
log-price correspondence is approximate; (ii) a controlled diagnostic of
parameter-free transforms, channel restrictions, and matched-capacity
encoders where each contrast changes one factor; and (iii) evidence on a
point-in-time CSI~300 five-minute panel that centering carries the main
effect, scale-only normalization does not help, the gain is mainly
price-channel driven and not explained by short-term reversal, and nuisance
removal is a larger and steadier effect than encoder choice.

\section{Problem Formulation}
\label{sec:problem}

Let $\mathcal{U}_t$ be the set of index constituents on trading day $t$,
determined point-in-time. For each instrument $u\in\mathcal{U}_t$ the input is
the window of $T$ preceding intraday bars over $F$ raw fields,
$\mathbf{x}_{u,t}\in\mathbb{R}^{T\times F}$, here $T=240$ five-minute bars
(five days of $48$) and $F=6$ (open, high, low, close, volume, traded
amount); no hand-crafted factors or cross-asset features are used. We write
$x_{\tau,f}$ for bar $\tau$ of field $f$ after training-only global
$\log(1+\cdot)$ preprocessing and standardization, dropping instrument and
day indices when clear.

The label is the next-day close-to-close return
$y_{u,t}=p^{\mathrm{c}}_{u,t+1}/p^{\mathrm{c}}_{u,t}-1$, with $p^{\mathrm{c}}$
the close price. A scoring model $s_\theta$ (an encoder followed by a linear
score head) maps each window to a scalar and is evaluated by the daily
cross-sectional Spearman correlation between scores and labels,
\begin{equation}
 \rho_t=\operatorname{corr}_{\mathrm{rank}}\!\bigl(
 \{s_\theta(\mathbf{x}_{u,t})\}_{u\in\mathcal{U}_t},
 \{y_{u,t}\}_{u\in\mathcal{U}_t}\bigr),
 \label{eq:rankic}
\end{equation}
averaged over test days (rank IC). Since $\rho_t$ depends only on
within-day ranks, a model is judged by the ordering it induces.

\textbf{Nuisance hypothesis.} For a raw price $p$ and a positive multiplier
$c$, the label $y$ is unchanged, but the transformed displacement is
proportional to $\log(1+cp)-\log(1+p)$, which approaches $\log c$ as $p$
grows but is not constant at low prices. The hypothesis therefore concerns
additive offsets in the transformed price channels: a per-instrument
constant added to $x_{\cdot,f}$ for a price field $f$ carries no ranking
information and should be removed. Mean removal is exactly invariant to this
perturbation even though the mapping from raw-price scaling is approximate.
The question is whether removing that offset, rather than rescaling
amplitudes or changing the encoder, explains the gains attributed to
instance normalization.

\section{Controlled Diagnostic Framework}
\label{sec:method}

\begin{figure*}[!t]
\centering
\includegraphics[width=\textwidth]{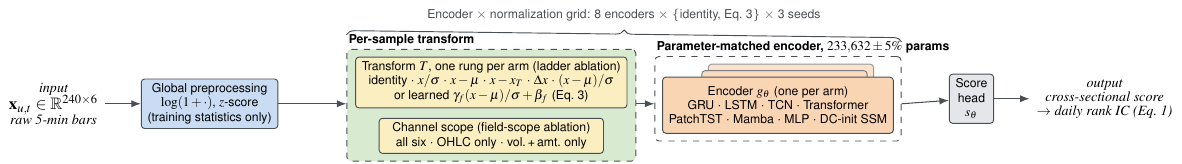}
\caption{Overview of the controlled diagnostic. Raw bars pass through
training-only global preprocessing, one per-sample transform (a ladder rung of
Sec.~\ref{ssec:ladder} or the learned Eq.~\eqref{eq:instnorm}, applied to all
six, price-only, or volume/amount channels), and one of eight
parameter-matched encoders with a score head; the encoder and head together
form $s_\theta$, and the scores are evaluated by daily rank IC. Exactly one
factor varies per contrast; every arm shares data, target, loss, optimizer,
and evaluator.}
\label{fig:overview}
\end{figure*}

Figure~\ref{fig:overview} shows the diagnostic pipeline: training-only
global preprocessing, one per-sample transform, and one of eight
parameter-matched encoders emitting a cross-sectional score. The pipeline is
held fixed while exactly one factor varies: the learnable normalization block
(Sec.~\ref{ssec:instnorm}), the ladder transform (Sec.~\ref{ssec:ladder}),
or its channel scope (Sec.~\ref{ssec:channels}); Secs.~\ref{ssec:protocol}
and \ref{ssec:stats} give the shared protocol and inference.

\subsection{Learnable instance normalization}
\label{ssec:instnorm}

For a window of length $T$, define
\begin{equation}
 \mu_f=T^{-1}\!\sum_\tau x_{\tau,f},\qquad
 \sigma_f^2=T^{-1}\!\sum_\tau(x_{\tau,f}-\mu_f)^2 .
 \label{eq:moments}
\end{equation}
The main comparison uses the normalization half of RevIN
\cite{kim2022revin},
\begin{equation}
 z_{\tau,f}=\gamma_f\frac{x_{\tau,f}-\mu_f}
 {\sqrt{\sigma_f^2+10^{-5}}}+\beta_f,
 \label{eq:instnorm}
\end{equation}
where $\gamma_f$ and $\beta_f$ are learned. RevIN's output denormalization
does not apply, since the output is a dimensionless ranking score. Instance
statistics come only from each sample's backward-looking window.

\subsection{Parameter-free transform ladder}
\label{ssec:ladder}

Equation~\eqref{eq:instnorm} removes offset and scale at once, so it cannot
say which of the two matters. The mechanism study therefore removes the affine
parameters and compares six zero-parameter transforms: identity; scale only
$x/\sigma$; centering $x-\mu$; last-value referencing $x-x_T$; first
difference $x_\tau-x_{\tau-1}$; and standardization $(x-\mu)/\sigma$.
Differences are formed between adjacent observed bars with a zero first
output. Centering, last-value referencing, and differencing all exactly
remove an additive transformed-input offset, although the latter two also
change the low-frequency response. Scale-only normalization leaves the
offset intact and is the control for amplitude effects. Because the ladder
adds no parameters, same-backbone arms are exactly comparable.

\subsection{Channel-restricted transforms}
\label{ssec:channels}

Each transform acts independently per sample and per field, so its scope can
be restricted to all six fields, to the four price fields only, or to volume
and traded amount only; unselected channels pass through unchanged. The
price-only scope tests the nuisance hypothesis directly; the volume/amount
scope is the control, since a gain persisting there would indicate generic
conditioning rather than price-offset removal.

\subsection{Matched-capacity controlled protocol}
\label{ssec:protocol}

All arms share cached tensors, target, loss, AdamW
\cite{loshchilov2019adamw}, one-cycle scheduling
\cite{smith2018superconvergence}, weight averaging, an 80-epoch budget,
patience 10, and evaluator. The shared loss combines Huber error, soft-rank
and Pearson correlations, top--bottom spread, and a dispersion penalty.
Batches remain within one day, so its ranking terms are cross-sectional.
Binary search matches the eight encoders to $233{,}632\pm5\%$ trainable
parameters. We compare GRU, LSTM, TCN, Transformer, and PatchTST
\cite{cho2014gru,hochreiter1997lstm,bai2018tcn,vaswani2017attention,nie2023patchtst},
Mamba \cite{gu2023mamba}, MLP, and a DC-initialized filter-bank SSM;
the SSM is our own controlled implementation, not a published pipeline.
Every arm has three matched seeds. The ladder and channel studies use two
fixed backbones: a linear stem (per-bar linear projection with mean and
last-bar readouts, no temporal convolution) and GRU.

\subsection{Statistical inference}
\label{ssec:stats}

For each contrast, we subtract daily IC within matched seed and average across
seeds; a length-10 moving-block bootstrap over dates
\cite{lahiri2003resampling} gives intervals that preserve short-range
dependence, conditional on the three seeds. Holm correction \cite{holm1979}
is applied within each comparison family, and TOST \cite{schuirmann1987tost}
with a pre-specified margin of $0.005$ rank IC keeps a non-significant
difference from being read as equivalence. As a robustness check, scores are
residualized on eight price--volume style proxies, including five-day
short-term reversal, before recomputing IC; industry dummies are unavailable
for this public panel.

\section{Experiments}
\label{sec:experiments}

\begin{figure*}[!t]
\centering
\includegraphics[width=0.74\textwidth]{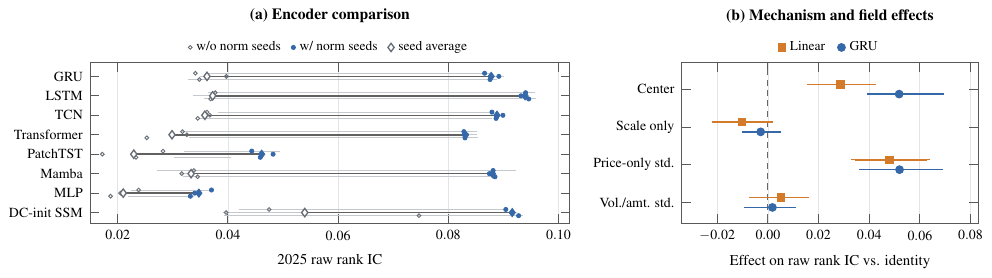}
\caption{Normalization effects. (a)~Per-seed mean raw rank IC (circles) and
three-seed averages (diamonds). (b)~Effects relative to the same-backbone
identity arm with matched-day 95\% block-bootstrap intervals.}
\label{fig:summary}
\end{figure*}

\subsection{Datasets and implementation details}
\label{ssec:data}

\textbf{Datasets.} We use CSI 300 five-minute bars from the public
Qlib archive \cite{yang2020qlib} (2019--2022) and Baostock (2023--2025).
Inputs use unadjusted prices, volumes, and traded amounts;
labels are back-adjusted close-to-close returns. Constituents
follow the latest Baostock membership snapshot effective
on or before each trading day. We restore Qlib prices and
volumes to unadjusted values, correct its 2020-09-28 per-segment
price rescaling before computing labels, and exclude
suspended or incomplete stock-days.

\textbf{Evaluation.} The temporal split is training in 2019--2023, validation
in 2024, and testing in 2025 (242 days; 2025-12-31 has no next-day label),
with a two-trading-day embargo. Global preprocessing statistics are fitted
on training data only. The primary metric is mean daily rank IC,
Eq.~\eqref{eq:rankic}, on raw scores; style-residualized IC is a robustness
check. The protocol of Sec.~\ref{ssec:protocol} yields 48
encoder-by-normalization, 36 ladder, and 24 channel-scope runs; contrasts and
the TOST margin were specified before inspecting the 2025 results.

\subsection{Main results: normalization across encoders}
\label{ssec:encoders}

\begin{table}[t]
\centering
\caption{Mean raw rank IC on the 242-day 2025 test period (three seeds).
$\Delta$ is the matched daily effect of instance normalization with a 95\%
block-bootstrap interval; all eight survive Holm correction. The left column
is our normalization ablation, not these methods as published.}
\label{tab:main}
\small
\setlength{\tabcolsep}{3.2pt}
\begin{tabular}{lccc}
\toprule
Encoder & w/o norm & w/ norm & $\Delta$ [95\% CI] \\
\midrule
GRU          & $0.0362$ & $0.0878$ & $+0.0516\ [0.0383,0.0670]$ \\
LSTM         & $0.0372$ & $0.0939$ & $+0.0567\ [0.0439,0.0717]$ \\
TCN          & $0.0358$ & $0.0888$ & $+0.0530\ [0.0383,0.0682]$ \\
Transformer  & $0.0299$ & $0.0830$ & $+0.0532\ [0.0390,0.0686]$ \\
PatchTST     & $0.0229$ & $0.0461$ & $+0.0232\ [0.0105,0.0368]$ \\
Mamba        & $0.0333$ & $0.0880$ & $+0.0547\ [0.0431,0.0674]$ \\
MLP          & $0.0210$ & $0.0347$ & $+0.0137\ [0.0020,0.0263]$ \\
DC-init SSM  & $0.0539$ & $0.0916$ & $+0.0376\ [0.0272,0.0505]$ \\
\bottomrule
\end{tabular}
\end{table}

Normalization raises rank IC for all eight encoders (Table~\ref{tab:main},
Fig.~\ref{fig:summary}(a)); all 24 seed-level contrasts are positive, and
all eight effects survive Holm
correction on raw IC, on style-residualized IC, and on IC residualized
additionally against two hand-crafted reversal factors (below). For the six
stronger encoders, normalization gains $0.0376$--$0.0567$, whereas their
normalized IC values span only $0.0109$ ($0.0830$--$0.0939$). None of their
15 pairwise differences survives Holm correction; the three pairwise
comparisons among GRU, TCN, and Mamba meet the $0.005$ TOST equivalence
criterion. The DC-initialized SSM,
whose zero-sum stem partially removes offsets by construction, starts highest
without normalization and gains least. PatchTST and MLP remain below this
group under the shared implementation. Averaged across the six
stronger encoders, the normalization gain is positive in every 2025 quarter
($0.0533$, $0.0803$, $0.0267$, and $0.0467$). On the 2024 selection year
the same seven gains recur ($+0.023$ to $+0.045$) while MLP shows none.
Daily IC of the normalized GRU has standard deviation $0.15$ (information
ratio $0.58$).

To test whether normalization merely exposes short-term reversal, we compute
two hand-crafted reversal factors on the same windows: the negated last price
relative to the window mean, and relative to the first bar. They reach raw IC
$0.055$ and $0.048$; each of the six stronger encoders exceeds the stronger
factor by $0.028$--$0.039$ (all Holm-significant), whereas PatchTST, MLP, and
the unnormalized GRU exceed neither, and residualizing scores on both factors
leaves all eight normalization effects significant at essentially unchanged
magnitude.

\subsection{Ablation: centering versus scaling}
\label{ssec:mechanism}

\begin{table}[t]
\centering
\caption{Parameter-free transforms: 2025 raw rank IC and change from the
same-backbone identity arm (parentheses).}
\label{tab:ladder}
\small
\setlength{\tabcolsep}{2.8pt}
\begin{tabular}{lcc}
\toprule
Transform & Linear & GRU \\
\midrule
Identity $x$               & $0.0369$ & $0.0362$ \\
Scale only $x/\sigma$      & $0.0266\ (-0.0103)$ & $0.0334\ (-0.0028)$ \\
Center $x-\mu$             & $0.0656\ (+0.0287)$ & $0.0880\ (+0.0518)$ \\
Last value $x-x_T$         & $\mathbf{0.0891}\ (+0.0522)$ & $0.0925\ (+0.0563)$ \\
Difference $\Delta x$      & $0.0721\ (+0.0352)$ & $\mathbf{0.0930}\ (+0.0568)$ \\
Standardize $(x-\mu)/\sigma$ & $0.0882\ (+0.0513)$ & $0.0879\ (+0.0517)$ \\
\bottomrule
\end{tabular}
\end{table}

Table~\ref{tab:ladder} separates the two halves of Eq.~\eqref{eq:instnorm}.
Centering gains $+0.0287$ on the linear stem and $+0.0518$ on GRU; both 95\%
intervals exclude zero, and these effects account for $56\%$ and $100\%$ of
each backbone's full-standardization gain. Scaling alone does not help:
$-0.0103$ on the linear stem and $-0.0028$ on GRU. Full standardization adds
$+0.0226$ over centering on the linear stem but nothing on GRU ($-0.0001$);
the reliable component is offset removal (Fig.~\ref{fig:summary}(b)).
The reference used to remove it also matters: last-value referencing and
differencing outperform simple centering in point estimates, but linear-stem differencing is unstable across seeds
($0.0987$, $0.0978$, $0.0198$), so the table establishes offset removal as the
common driver without isolating a low-frequency advantage beyond it.

\subsection{Ablation: field-specific effects}
\label{ssec:fields}

\begin{table}[t]
\centering
\caption{All-field and restricted transforms on 2025 raw rank IC. Price denotes
OHLC; vol./amt. denotes volume/traded amount. Identity is $0.0369$ (linear) and
$0.0362$ (GRU).}
\label{tab:fields}
\small
\setlength{\tabcolsep}{4.6pt}
\begin{tabular}{llccc}
\toprule
Backbone & Transform & All & Price & Vol./amt. \\
\midrule
Linear & Center      & $0.0656$ & $0.0569$ & $0.0427$ \\
Linear & Standardize & $0.0882$ & $0.0848$ & $0.0419$ \\
GRU    & Center      & $0.0880$ & $0.0921$ & $0.0402$ \\
GRU    & Standardize & $0.0879$ & $0.0882$ & $0.0380$ \\
\bottomrule
\end{tabular}
\end{table}

Table~\ref{tab:fields} locates the effect. Price-only standardization retains
$93\%$ of the linear-stem gain and $101\%$ of the GRU gain. Its advantage over
the volume/amount arm is $+0.0428$ on the linear stem and $+0.0502$ on GRU;
both intervals exclude zero after Holm correction, while the volume/amount
arm gains only $+0.0051$ and $+0.0018$, neither distinguishable from zero.
Centering alone reproduces this separation on GRU ($+0.0519$,
Holm-significant) and a smaller one on the linear stem ($+0.0142$).
The effect is therefore predominantly price-channel
driven, as Sec.~\ref{sec:problem} predicts; the data do not
show that volume and amount are irrelevant.

\section{Conclusion}
\label{sec:conclusion}

We formulated price level as an additive-offset nuisance in the transformed
input space of cross-sectional return ranking and built a controlled
diagnostic to test whether removing it explains the gains attributed to
instance normalization. Normalization helps all eight parameter-matched
encoders, and among the six stronger ones its gains ($0.0376$--$0.0567$)
exceed the $0.0109$ spread between encoders. Centering carries the main
effect while scale-only
normalization does not help, so instance normalization is not one
indivisible operation; the gain is predominantly price-channel driven and
is not explained by known short-term reversal. Comparisons that normalize
only one model therefore risk crediting an input-representation gain to
architecture. Nevertheless, encoder choice still matters: PatchTST and MLP
remain below the other six encoders after normalization under the shared
protocol. The practical lesson: center transformed price channels
before crediting performance differences to the encoder, and test scale
removal separately. The evidence covers one market, bar frequency, and
test year with three seeds.

\clearpage
\section{Acknowledgments}
No funding was received for conducting this study. The authors have no
relevant financial or nonfinancial interests to disclose.

\begingroup\ninept\normalsize
\bibliographystyle{IEEEbib}
\bibliography{refs}
\endgroup

\end{document}